\documentclass[final,10pt,twocolumn]{elsarticle}

\usepackage[switch]{lineno}
\usepackage{hyperref}
\usepackage{multirow}
\usepackage{amsmath,amsfonts,amssymb,amscd,amsthm,xspace}
\usepackage{graphicx}
\usepackage{epstopdf}
\usepackage{booktabs}
\usepackage[shortlabels]{enumitem}
\usepackage[usenames,dvipsnames,svgnames,table]{xcolor}
\usepackage{rotating}
\usepackage{listings}
\usepackage{longtable}
\modulolinenumbers[10]

\journal{Nuclear Instrument and Methods in Physics Research A}

\begin{document}

\begin{frontmatter}

\title{Fluid-dynamics in the Borexino Neutrino Detector: behavior of a pseudo-stably-stratified, near-equilibrium closed system under asymmetrical, changing boundary conditions}

\author[myamericanaddress,myitalianaddress]{D. Bravo-Bergu\~no}

\author[RiccardoAddress]{Riccardo Mereu}
\author[myamericanaddress]{R.B. Vogelaar}
\author[RiccardoAddress]{F. Inzoli}

\address[myamericanaddress]{Physics Department, Virginia Tech, 24061 Blacksburg, VA (USA)}
\address[myitalianaddress]{INFN Sezione Milano, Via Celoria 16, 20133 Milano (Italy) - (+39) 3394636849 - \textit{david.bravo@mi.infn.it}}
\address[RiccardoAddress]{Department of Energy - Politecnico di Milano, via Lambruschini 4, 20156 - Milano, Italy}

\begin{abstract}
The strategy to install Borexino's Thermal Monitoring and Management System (BTMMS) successfully stabilized the thermal environment inside the Borexino neutrino observatory, which is understood to be a necessary step to improve and minimize radioactive background contamination inside the active volume of the detector, allowing for it to achieve better sensitivity in the regions of interest. Two-dimensional numerical simulations to achieve a proper understanding of Borexino's fluid-dynamics were developed and optimized for different regions and periods of interest, focusing on the most critical effects that were identified as influencing background concentrations. Literature experimental case studies were reproduced to benchmark the method and settings, and a Borexino-specific benchmark was constructed in order to validate the model's thermal transport. Finally, fully-convective models were implemented to understand general and specific fluid motions impacting the active detector volume.
\end{abstract}

\begin{keyword}
Neutrino detector \sep
Computational Fluid Dynamics \sep
Thermal control \sep
Radiopurity \sep
Background stabilityÊ\sep
Natural convection
\end{keyword}

\end{frontmatter}

\linenumbers
\begin{nolinenumbers}

\section{Introduction}
\label{sec:intro}

The Borexino liquid scintillator detector is devoted to performing high-precision neutrino observations. In particular, it is optimized to study the low energy part of the solar neutrino spectrum in the sub-MeV region, having the precision measurement of the $^7$Be solar neutrinos as its design objective. Borexino has succeeded in performing high-precision measurements of all the major components of the solar neutrino spectrum (first direct detections of \textit{pp}\cite{pp}, \textit{pep}\cite{pep}, $^7$Be\cite{7Be}, and lowest (3 MeV) threshold observation of $^8$B\cite{8B}), as well as in reaching the best available limit in the subdominant CNO solar neutrino rate\cite{8B}, with just the DAQ time of 767 days comprising its first dataset \textit{Phase 1} from 2007-10. Geoneutrinos have also been detected by Borexino with high significance (5.9$\sigma$\cite{geo}) thanks to the extremely clean $\overline{\nu}$ channel. Results on searches for new particles, (anti)neutrino sources and rare processes like \cite{sterile_old}, \cite{antinu_sources}, \cite{pauli_trans}, \cite{e_decay}, \cite{axions} are expected to gain even more relevance during the SOX phase of the experiment, where a $\overline{\nu}_e$ generator will be placed in close proximity to the detector, in order to probe for anomalous oscillatory behaviors and unambiguously cover the allowed phase space for light sterile neutrinos\cite{SOX}.

These results were possible thanks to the unprecedented, extremely radio-pure conditions reached in the active section of the detector --achieved thanks to a combination of ultra-clean construction and fluid-handling techniques as well as dedicated scintillator purification campaigns\cite{purif}. Detailed detector response determination was made possible thanks to very successful internal calibration campaigns\cite{calib} which did not disturb the uniquely radio-pure environment. Moreover, results with even higher precision are under development thanks to the \textit{Phase 2} dataset, started in late 2011, which offers greatly enlarged statistics with improved background conditions, being parsed with new analysis techniques. 

Reduced background levels in the \textit{Phase 2} dataset have raised the need for increased stability in their spatio-temporal distribution inside the detector, due to the low statistics available for determining their rate, particularly for some background species. The liquid nature of the scintillator in use by Borexino means the best strategy to ensure background stability is to minimize fluid mixing, namely by means of external environmental control and stabilization. It is assumed the extremely dilute concentrations of background radioisotopes are carried by the fluid movement in ideally point-like, non-interacting particulates --thereby establishing a direct correlation between fluid dynamics and background migration, which is only attenuated through the corresponding radioisotopes' lifetimes.

Section 2 of this paper will detail the recent background situation in Borexino and the correlation existent with temperature trends within the detector, pointing toward the strategy chosen to attempt an improvement in Fiducial Volume (FV) background levels. Section 3 will discuss the benchmarking strategy for the Computational Fluid Dynamics (CFD) convective simulations that would confer confidence on the simulative approach used, as well as the empirical LTPS-data-based thermal benchmarking performed for the Borexino geometry. Section 3 will instead focus on the Inner Detector models developed in order to understand fluid movement inside this closed, near-equilibrium system. It will also showcase the evolution toward a more focused model with greater detail around Borexino's active section, the Inner Volume (IV). Finally, Section 4 will discuss the conclusions reached through these models, and the perspectives on new studies building upon the present work.

\section{Background stability and CFD}
\label{sec:stability}

Borexino, located in the Hall C of the Gran Sasso National Laboratories' (LNGS) underground facilities (3,800 m w.e.), measures solar neutrinos via their interactions with a 278 tonnes target of organic liquid scintillator. For more details about its structure, design and signal/background issues relevant to the present discussion, as well as the Thermal Monitoring and Management System developed and installed during 2014-16, refer to \cite{previous_paper}.

The installation of the Latitudinal Temperature Probe System (LTPS) multi-sensor hardware, especially the internal Phase I, offered an unprecedented chance at utilizing abundant data from evenly distributed points on both sides of the detector for applications beyond trivial temperature monitoring, such as the thermal transport constant from the Outer to the Inner detector, liquid stratification, 180$^{\circ}$-resolution thermal asymmetries... Conductive 2D simulations were also developed with the aim of establishing a first layer of information about which heat transfer processes were conductive-dominated, in particular the cooldown effect dominated by the bottom heat sink in contact with the local aquifer temperature, the cooling constant and behavior it would cause, the extent of the Thermal Insulation System (TIS) insulating power and expectable boundary temperature trends, and the influence of structures as temperature bridges, including the conductive heat transmission expectable from the Active Gradient Stabilization System (AGSS) heating circuit. This would provide information needed to validate the simulative approach and move toward more detailed simulations.

However, as is obvious, these conductive cases would not account for convection or fluid movement, imposing a clear limitation in simulating thermal trends away from stable stratifications or solid structures, as well as understanding where backgrounds would be led by fluid-dynamical currents.

Considering the internal LTPS dataset, we can neglect the detailed modeling of the water convection in the WT: a convective model of the full detector, even in 2D, would either be too coarse for the expected convective speeds to surface from under the numerical noise, or take up too much computing time (\textit{i.e.} the simulated time of interest, which is on the order of months, would approach the real time, negating its usefulness as a predictive tool, and delaying too much the results' availability). 

\section{CFD Methodology}
\label{sec:methodology}

In order to validate the CFD approach \cite{riccardo} in reproducing the main characteristics and fundamental phenomena inside Borexino, such as closed system, Newtonian water-like fluids, prevalent natural convection, vertical temperature difference ($\sim$10$^{\circ}$C) and thermal stratification, several benchmarking cases have been modeled and related results compared with available experimental data.

\subsection{Design parameters}

The choice of dimensions and $\Delta T$ for the following benchmark models needed to be motivated to at least lie close to, or ideally overlap, Borexino's regime of interest. The determination of the Rayleigh number for Borexino offers the simplest, most rigorous way of relating seemingly dissimilar geometries to the detector case. The definition of the Rayleigh number is very dependent on the model geometry, and in non-standard ones (such as Borexino's spheric geometry with distributed, gradual temperature differences) may be somewhat arbitrary if not keeping a close watch on the phenomenon under study. The Rayleigh number ($Ra$) is a dimensionless parameter defined, in general, as:

\begin{equation}
Ra = \frac{\beta [K^{-1}] \Delta T [K] g [m/s^2] L^3 [m^3]}{\nu ^2 [m^4/s^2]} Pr
\label{eq:Rayleigh}
\end{equation}

where $\beta$ is the thermal expansion coefficient of the fluid, $\Delta T$ is the temperature difference in the characteristic lengthscale of the system, $g$ is the gravitational field acting on the system, $L$ is the characteristic lengthscale for natural convection in the system, $\nu$ is the kinematic viscosity of the fluid and $Pr$ is the Prandtl number, which is itself defined as the quotient between the momentum diffusivity and the thermal diffusivity. In practice, the Prandtl number is only dependent on the fluid's nature and state. The quotient multiplying $Pr$ is referred to as the \textit{Grashof number} $Gr$, which is a measure between the buoyancy and viscosity forces on a fluid.

As can be inferred from this definition, the Rayleigh number is most dependent on the characteristic lengthscale $L$ for convection in the considered system. $Ra$ is therefore a way of relating buoyancy-driven fluid flows coming from different fluid natures, conditions and system geometries --and therefore, contains information about the convective/conductive dynamics of a fluid flow, irrespective of the fluid. This is in contrast to the Grashof number, which depends upon the fluid under consideration.

If we consider the typical overall gradient of Borexino's IV to be $\sim$5$^{\circ}$C, over the 8.5 m between the top and bottom poles of the vessel, we get $\sim$1.7 m/$^{\circ}$C: that is, $\sim$17 cm separating each 0.1$^{\circ}$C isotherm. Considering this is our $L$, $Ra \sim \mathcal{O}(10^7-10^8)$ (with $Pr$=7.78 for PC, $\beta _{PC}^{10C} \sim$10$^{-3}$ K$^{-1}$, and $\nu_{PC}^{10C} \sim$7$\cdot$10$^{-7}$ m$^2$/s). We consider the $\mathcal{O}(0.1)^{\circ}$C temperature differences routinely happening in short timescales in Borexino, which may be causing the internal stirring concerning us. If the overall gradient was very large, the isotherms would be very close together, and a given $\Delta T$ seeping in from the outside would show up at a smaller lengthscale than if the overall gradient was smaller, and the isotherms were farther apart from each other --in which case the isotherm displacement to match the boundary condition would occur over larger lengthscales.

We are of course assuming a linearly-stratified fluid, which is not the real case in Borexino (which exhibits a laxer stratification on the top than on the bottom). Therefore, we should keep in mind the order-of-magnitude Rayleigh number estimate above would be approximately 1-2 order(s) of magnitude larger, locally, on the top, and smaller on the bottom. Consequently, we can estimate \textit{Borexino's Rayleigh range} as $Ra$ $\epsilon [\mathcal{O}(10^6),\mathcal{O}(10^9)]$.

\subsection{Governing Equations}
\label{governingequations}

The finite volume commercial solver \textit{ANSYS-Fluent} is used for modeling the flow field via mass, momentum and energy conservation equations for incompressible Newtonian fluids with constant viscosity and density. Governing equations of mass, momentum and energy are numerically treated using the Direct Numerical Simulation (\textit{DNS}) approach for laminar flows, as reported in Eq. \ref{mass}, \ref{eq:momentum} and \ref{energy}, respectively.

\begin{equation} \label{mass}
\frac{\partial\rho}{\partial t}+\nabla\cdot\left(\rho\mathbf{u}\right)=0
\end{equation}

\begin{equation} \label{eq:momentum}
\frac{\partial{\rho\mathbf{u}}}{\partial t}+\nabla\cdot(\rho\mathbf{uu})=-\nabla p+\nabla\cdot\bar{\bar{\tau}}+\bar{\rho}{\mathbf{g}}
\end{equation}

\begin{equation} \label{energy}
\frac{\partial \rho E}{\partial t}+\nabla\cdot(\mathbf{u}(\rho E+p))=\nabla\cdot(k\nabla T+(\bar{\bar{\tau}}\cdot\mathbf{u}))
\end{equation}

where \textit{\textbf{u}}, \textit{p} and \textit{E} represent the velocity vector, static pressure and energy, respectively, $\bar{\rho}\mathbf{g}$ the gravitational body force, $\kappa$ the thermal conductivity and $\bar{\bar{\tau}}$ the stress tensor \citep{Malalasekera2007}. For the natural-convection flows the fluid density is modeled as a function of temperature. the Boussinesq model used here treats density as a constant value in all solved equations, except for the buoyancy term in the momentum equation, where it is approximated as:

\begin{equation} \label{boussinesq}
\bar{\rho}\mathbf{g}=(\rho-\rho_{0})\mathbf{g}\simeq -\rho_{0}\beta(T-T_{0})\mathbf{g}
\end{equation}

where $\beta=-\frac{1}{\rho}(\frac{\partial \rho}{\partial T})_p$ is the thermal expansion coefficient.

\subsection{Geometrical modeling}

The numerical domain has been defined for each benchmarking case and Borexino reproducing the real geometry with a two-dimension full-scale model. The computational mesh used to discretize the domain is a structured Cartesian or unstructured polygonal/polyhedral grid, depending on the geometry. Specific refinements near the walls are applied to take into account the viscous and thermal boundary layer.

The mesh size ($\Delta x$) is defined for each modeled geometry and it is based on a preliminary mesh sensitivity analysis. This permitted to quantify the influence of different grid sizes selecting the largest acceptable size, with computational grids ranging from $\mathcal{O}(10^{4})$ to $\mathcal{O}(10^{5})$ elements.

\subsection{Numerical modeling}
\label{sec:Numerical modeling}

All the domain modeled here are closed and boundary conditions are imposed considering no-slip conditions for the fluid dynamics and adiabatic or fixed external temperature for heat transfer conditions.

The solver used for performing the transient simulations is based on the coupling pressure-velocity PISO algorithm. It is able to guarantee the convergence at each time step, through inner loops, using a restrictive Courant-Friedrichs-Lewy (\textit{CFL}) condition ($CFL\leq1$) to maintain the necessary accuracy. The time-step size $\Delta t$ is defined considering the physical $\Delta T_{p}$ and numerical Fourier stability analysis $\Delta T_{Fo}$ natural convection constrains:

\begin{equation} \label{physical}
\Delta t_{p} = \frac{\tau}{4} \approx \frac{L}{4\sqrt{\beta g L \Delta T}}
\end{equation}

\begin{equation} \label{numerical}
\Delta t_{Fo} = \frac{Fo(\Delta x)^{2}}{\alpha}
\end{equation}

with $\tau$ the time constant, $L$ the characteristic length, $Fo$ the Fourier number (limited to $Fo=0.1$), $\Delta x$ the grid size and $\alpha=\frac{k}{\rho c_{p}}$ the numerical diffusivity. Based on fluids properties, operational conditions and geometrical characteristics the maximum time-step size has been defined for each simulation presented below.

The same discretization schemes have been used for all simulations, as reported in Table \ref{tab:numerical schemes}.

\begin{table}[h]
\begin{centering}
\begin{tabular}{|c|c|}
\hline
\textbf{Term} & \textbf{Scheme}\tabularnewline
\hline
\hline
Transient & First order implicit\tabularnewline
\hline
Gradient & Least Squares Cell Based\tabularnewline
\hline
Pressure & Body Force Weighted\tabularnewline
\hline
Momentum, Energy & Third order MUSCL\tabularnewline
\hline
\end{tabular}
\par\end{centering}
\caption{Numerical Schemes \label{tab:numerical schemes}}
\end{table}

\section{Benchmarking validation}
\label{sec:benchmarking}

\subsection{Phenomenological benchmarks}
\label{sec:phenobench}

The level of currents that represent a concern inside Borexino's IV is derived from the background's half-life ($\tau_{1/2}^{^{210}Po} \sim$ 138 days) and the IV's dimensions (4.25 m nominal radius, or $\sim$1 m from the vessel where the $^{210}$Pb progenitor sits and provides an "inexhaustible" $^{210}$Po source for our purposes). This means that currents under $\mathcal{O}$(10$^{-7}$) m/s would be so slow that more than half of the detached polonium will decay away in the trip, even under directly radial motion. Therefore, the level of admissible numerically-induced systematic uncertainties should not exceed this magnitude and ideally be $\leq \mathcal{O}$(10$^{-8}$) m/s.

Simple scenarios involving a cylindrical 2D geometry were implemented to characterize the basic phenomena at work in a well-studied scenario with a regular square mesh that ideally avoids the creation of preferential directions. This rectangular mesh grid employed features an average cell size of $\sim$3 cm (11 cm$^2$). Initial conditions for all cases were set as a linearly-stratified temperature gradient of [10,18]$^{\circ}$C according to:

\begin{equation}
T(h)=T_2 + (T_1-T_2) \frac{h-h_0}{H}
\end{equation}

where H is the cylinder's height (13.7 m, to keep it within Borexino's dimensions, along with its width of 11.2 m) and $T_1$ ($T_2$) is the top (bottom) temperature.

A scenario where no motion would be expected established the level of irreducible numerical noise for the model at $<$3.5$\cdot$10$^{-7}$, with an undefined, random pattern across the model.

Sudden temperature variations on the bottom (raising $T$) and/or top (reduced $T$) surfaces, keeping the walls with an adiabatic boundary condition, showed regional effects circumscribed to those areas, which extended only until the height of the corresponding interior isotherm was reached. Threshold for convection triggering recirculation cell formation was $\Delta T >$0.1$^{\circ}$C. Equivalent dynamics were found for both top and bottom. Largest achieved currents were $\mathcal{O}$(dm/s). This result proves the inherently-safe principle of operation for the Active Gradient Stabilization System based on heat application on Borexino's top dome\cite{other_paper}.

In contrast, the application of non-adiabatic boundary conditions to all boundary surfaces, including the walls, prompted the generation of a global convective mode spanning the whole cylinder geometry, showing varied characteristics depending on the $\Delta T$, but organized around robust currents of rising/falling fluid along the walls, and weaker recirculation currents along the central axis, with the addition of meandering horizontal currents or recirculation cells for large $\Delta T$. Uniform, height-weighted or delayed (through the addition of varying thicknesses of insulation on the model's boundaries) $\Delta T$s were employed to characterize the different behaviors --nevertheless, an important final conclusion is that there is no "allowable" threshold on the amount of temperature difference that would not induce a global circulation pattern, if $\Delta T/\Delta t$ is small. Some currents enter the realm of the resolution limit for the model ($\mathcal{O}$(10$^{-9}$-10$^{-8}$ m/s), and would not be relevant in Borexino's case, but the organized convective structure remains in place.

\paragraph{Literature models and conditions} Reproduction of experimental results from selected literature examples were identified as the proper benchmarking strategy. The selected benchmarks \cite{Yin, Garg, Goldstein} are characterized by two-dimensional geometry, with circular cylindrical section (or spherical, but it is identical in the two-dimensional case) and the presence of a concentric annulus fully contained inside, with different ratios between the inner and outer radii, as shown in Table~\ref{table:benchmark_conditions}. The inner annulus' outer surface is the one at high temperature and the exterior's inner one at low temperature for all cases presented here. The analysis focuses on the fluid behavior in the space between and it is initially considered at a constant, volume-weighted, mean temperature defined by:

\begin{equation}
T_m = \frac{(r_{av}^3-r_i^3)T_i + (r_o^3-r_{av}^3) T_o}{r_o^3-r_i^3}
\end{equation}

where $r_{av}$ is the average radius $(r_o+r_i)/2$ and $r_i$ ($T_i$), $r_o$ ($T_o$) are the inner and outer radii (temperatures), respectively\cite{Yin}.

The numerical geometry for the benchmarking cases is a 2D model of the real geometry, with a structured mesh including a local refinement close to the walls. In all the geometries the minimum cell size $(\Delta x)_{min}$ ranges from $2\cdotp10^{-4}$ to $5\cdotp10^{-5}$ m and the number of cells from $2.3\cdotp10^{4}$ to $3.68\cdotp10^{5}$ for coarse and fine mesh, respectively. Numerical methods and algorithms are those described in Section \ref{Numerical modeling}, with a time-step size$\Delta t$ ranging from $2$ to $9$ s for fine and coarse mesh, respectively

For the first cases, taken from\cite{Yin}, the Grashof numbers given in the reference were converted to the dimensionless Rayleigh number and this was taken as reference to calculate the inner/outer surface temperatures, as well as the fluid's parameters according to Equation~\ref{eq:Rayleigh}. This involved an amount of (informed) arbitrariness, since the literature reference did not indicate the absolute temperatures they worked with. For that reason, ranges around Borexino's 10-20$^{\circ}$C were chosen when possible. The fluid employed was water, since the reference parameters are much better constrained than for PC/benzene at the small $\Delta T$s involved. Sometimes, owing to the low $Ra$ used, the temperature difference for a high-viscosity fluid such as water would be too small ($<\mathcal{O}(10^{-3}\ ^{\circ}$C)), and air was chosen instead. Dimensions were kept as in the reference. 

\begin{table*}[t]
\centering
\begin{tabular}{p{1.5cm} p{1cm} c c c c}
\multicolumn{2}{c}{\textbf{Medium}} & \multicolumn{4}{c}{\textbf{Characteristics}} \\
\cmidrule(r){1-2}
\cmidrule(r){3-6}
\textbf{Ref} & \textbf{Model} & \textbf{Ra} & \textbf{$r_o/r_i$ } & \textbf{$\Delta T$} (K) & \textbf{Features of interest} \\
\cline{1-6}
Air\cite{Yin} & Air & 5880 & 1.78  & 0.64 & Chimney \\
Air\cite{Yin} & Air & 5880 & 1.4 & 2.33 & Chimney + upper cells\\
Water\cite{Garg} & Water & 90000 & 2 & 0.24 & Isotherm and circulation pattern \\
Air\cite{Garg} & Water & 250000 & 2 & 0.11 & Isotherm and circulation pattern (kidney-shape) \\
Air\cite{Yin} & Air & 739200 & 2.17 & 0.02 & Kidney-shaped cells + elongated cells \\
Water\cite{Garg} & Water & 10$^6$ & 2 & 0.09 & Square-kidney pattern + upper and lower vortices \\
Water\cite{Yin} & Water & 1.5$\cdot$10$^6$ & 2.17 & 0.06 & Circulation pattern + vortices\\
Water\cite{Goldstein} & Air & 2.51$\cdot$10$^6$ & 2.6 & 9.26 & Chimney, vortex, flat isotherms \\
Water\cite{Yin} & Water & 7.1$\cdot$10$^6$ & 1.78 & 0.33 & Distortion of steady flow into unsteady vorticity \\
Water\cite{Yin} & Water & 10$^7$ & 1.78 & 2.3 & Small circulation features in medium-scale\\
 & & & & & pattern at change of regime: double vortex, \\
 & & & & & shear structure, chimney unsteadiness. \\
Water\cite{Yin} & Water & 21.6$\cdot$10$^6$ & 2.17 & 0.19 & Upper cell + detachment of outward current \\
\end{tabular}
\caption{Summary of the operating conditions for the different annuli benchmarking simulations. The ratio between external and internal radii is $r_o/r_i$, and $\Delta T$ is the temperature difference imposed between the inner (hotter) and outer (colder) annuli.}
\label{table:benchmark_conditions}
\end{table*}

\paragraph{Results}

A summary of the achieved results, based on the main relevant features in each of the benchmark cases, is included in Table~\ref{table:benchmark_summary}.

\begin{table*}[t]
\centering
\begin{tabular}{p{1.5cm} p{1cm} c c c c}
\multicolumn{2}{c}{\textbf{Medium}} &  & \multicolumn{3}{c}{\textbf{Features}} \\
\cmidrule(r){1-2}
\cmidrule(r){4-6}
\textbf{Ref} & \textbf{Sim} & \textbf{Ra} & \textbf{Small} & \textbf{Medium} & \textbf{Large} \\
\cline{1-6}
Air\cite{Yin} & Air & 5880 & - & \textbf{Chimney} & \textbf{Crescent} \\
Air\cite{Yin} & Air & 5880 & Upper vortices & \textit{Flow direction} & \textbf{Crescent} \\
Water\cite{Garg} & Water & 90000 & - & \textit{Cell center} & \textbf{Crescent, isotherms} \\
Air\cite{Garg} & Water & 250000 & - & \textit{Cell center} & \textbf{Kidney, isotherms} \\
Air\cite{Yin} & Air & 739200 &  \textit{Vortex structure} &  \textit{Cell center} & \textbf{Kidney, isotherms} \\
Water\cite{Garg} & Water & 10$^6$ & \textit{Vortices} & \textit{Streamlines} & \textbf{Isotherms, fluid flow} \\
Water\cite{Yin} & Water & 1.5$\cdot$10$^6$ & Stronger vortices & \textit{Upper flow} & \textbf{Fluid flow} \\
Water\cite{Goldstein} & Air & 2.51$\cdot$10$^6$ & \textbf{Heat transfer} & \textit{Isotherms} & \textbf{Nusselt} \\
 & & & \textbf{Nusselt} & &  \\
Water\cite{Yin} & Water & 7.1$\cdot$10$^6$ & Stronger vortices & \textit{Transition threshold} & \textbf{Fluid flow} \\
Water\cite{Yin} & Water & 10$^7$ & \textit{Double vortex} & \textbf{L-shape} & \textbf{Fluid flow}  \\
 & & & \textit{and shear} & \textbf{detached features} & \textbf{and structure} \\
Water\cite{Yin} & Water & 21.6$\cdot$10$^6$ & \textit{Vortex position} & \textbf{Vortex} & \textbf{Flow} \\
 & & &  & \textbf{detached features} & \textbf{stagnant region} \\
\end{tabular}
\caption{Summary of CFD literature benchmarking results, showing well-reproduced features (\textbf{boldface}), features present with small deviations from literature (\textit{itallics}) and features which are absent or present with large deviations (normal font).}
\label{table:benchmark_summary}
\end{table*}

We can confront the average Nusselt number $\overline{Nu}$ from the inner cylinder's exchanged power (59.654 W). The Rayleigh number is the adimensional number to compare natural convection in different cases while the Nusselt number compares the heat exchange behavior of different cases. From the reference, we know that:

\begin{equation}
\overline{Nu}_{conv} = \frac{\overline{h}_i D_i}{\kappa}
\label{eq:nusselt}
\end{equation}

where $\overline{h}_i$ is the local heat transfer coefficient in the inner cylinder, $D_i$ is its diameter and $\kappa$ is the thermal conductivity. We also know that $Nu = \overline{\kappa}_{eq} \cdot Nu_{cond}$, and since $Nu_{cond} = 2 / ln(D_o/D_i)$=2.09, we can calculate the $Nu$ from the $\overline{\kappa}_{eq}$=7.88 provided in the reference's Table 1 (for our $Ra$=2.51$\cdot$10$^6$). Indeed, the total average $Nu$ is then $Nu$=7.88$\cdot$2.09=16.47.

We shall compare this adimensional number to the one obtained through the data from the simulation, because the Nusselt number offers a way to find a correspondence between very different cases, from the point of view of geometry and fluids, with regard to heat exchange. It is noted the heat transfer coefficient cannot be the same as in the reference case because we use water instead of air for reference fluid, among other model considerations. However, the fact that the model is not precisely equal to that in the paper increases confidence in the validity of the modelization. 

In any case, from the numerical point of view, we have a different $\overline{\kappa}_{eq}$, and we want to obtain the $\overline{Nu}$ shown in Equation~\ref{eq:nusselt}, so we need $\overline{h}_i$, defined as:

\begin{equation}
\overline{h}_i = \frac{Q}{\pi D_i Z (T_i-T_o)}
\end{equation}

where $Q$ is the total power exchanged and $Z$ is the depth of the cylinder (20.8 cm). The $\Delta T$ is 9.26$^{\circ}$C. With this data, we have $\overline{h}_i$=276. This yields a $\overline{Nu}$=16.431 (with a thermal conductivity $\kappa$=0.6 W/(m$\cdot$K)), which is in very good agreement with the $Nu$ found with the reference data.

We can also compare the $h$ for different points along the inner/outer cylinder's surface to establish a comparison with Figure 8 in the reference, which shows the local heat transfer coefficient versus the angular position considered, from the total surface heat flux (W/m$^2$) shown in Figure~\ref{fig:Goldstein_heattransfer}. Values were converted to Nusselt number in order to establish a one-to-one comparison irrespective of differences in the employed fluid.

\begin{figure}[tb]
\centering 
\includegraphics[width=1\columnwidth]{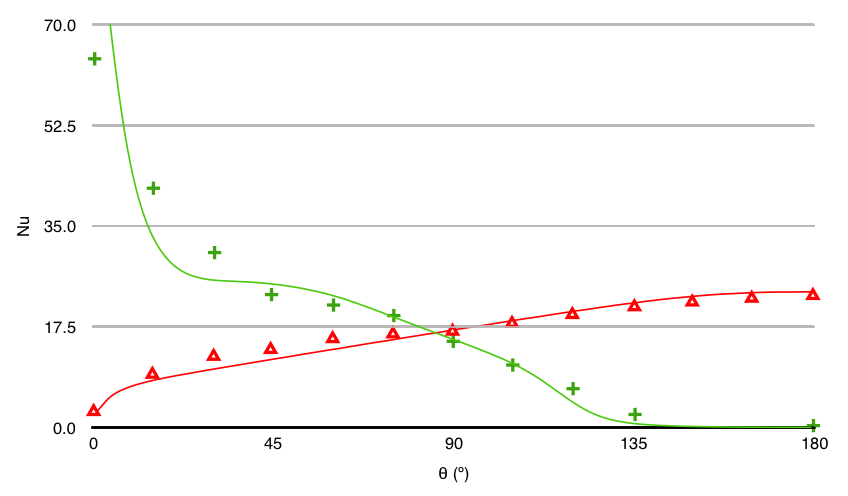} 
\caption{Comparison of the adimensional local Nussel numbers, between this paper's CFD results (solid lines) and the reference's\cite{Goldstein} (data points), for the inner (red) and outer (green) annulus surfaces.}
\label{fig:Goldstein_heattransfer} 
\end{figure}

From the values reported, we can readily see the extremal point for the inner cylinder at 0$^{\circ}$ (top) tends to zero, as is the case for the simulated model. This is expectable since it will be in this region where the "chimney" structure will rise, at a very similar temperature to the imposed inner cylinder one. This tendency starts at a steep decline $\sim$4 on the $\kappa_{eq}$ (i.e. 4$\cdot$2.09=8.36) on the reference plot, which is in reasonable agreement with the value we get from the $\sim$1000 W/m$^2$ from the \textit{FLUENT} plot, from the formula $h_i = q_{w_i}/ \Delta T$ and converting that to a Nusselt number by dividing by the thermal conductivity $\kappa$=0.6 W/(m$\cdot$K), into $Nu$=6.4. On the other extremal point at 180$^{\circ}$, which in the reference Figure 8 lies at $\kappa_{eq}\sim$11 for the inner cylinder, here we find a value of $\sim$3.7$\cdot$10$^3$W/m$^2$. This translates to a Nusselt of $Nu$=23.7, in even better agreement with the plot-provided value of $Nu$=11$\cdot$2.09$\approx$23.

For the outer cylinder, we can do similar calculations to compare the extremal point at 0$^{\circ}$ between the $\kappa_{eq} \sim$31 in the reference plot ($Nu\approx$64.8) and the 4500 W/m$^2$ from the surface heat flux plot ($Nu$=74.9), or the plateau value at $\sim$60$^{\circ}$ of around 11 in the reference plot ($Nu \approx$23) and the $\sim$1500 W/m$^2$ in the surface heat flux plot ($Nu$=24.97). The curves also follow similar trends, as can be appreciated in Figure~\ref{fig:Goldstein_heattransfer}.

In conclusion, the benchmarking showed good reproducibility of the thermal environment (when available to compare in the references) as well as the large- and medium-scale features present in each of the cases. Furthermore, while some small-scale features were not well-reproduced in some of the lowest Rayleigh number cases (and thus further from Borexino's regime), the general fluid flow pattern was faithfully reproduced in practically all regions of all cases.

\subsection{Borexino thermal benchmark} 

It is advisable to benchmark not only the general reproducibility of results just described, but also the model behavior in our particular cases of interest. Although we have no way of directly measuring fluid-dynamic effects inside the SSS, apart from the limited inference obtained from the background movement analyses, we do have a good thermal transport probing system: the Phase I LTPS sensors. Indeed, we were able to empirically measure the time constant for the thermal inertia between the inside and outside of the SSS from the Phase I.a and I.b sensors\cite{previous_paper}. We can now employ the temperatures registered on the outside (\textit{i.e.} in the water around the SSS, through the Phase I.b probes) and study their transmission toward the inside of the Sphere. The most interesting feature here is that the simulated transmitted temperatures for the inside of the SSS can be confronted with the internal recorded temperatures (\textit{i.e.} in the buffer just inside the SSS, through the Phase I.a probes), and therefore establish the level of fidelity on thermal transport that the CFD strategy can offer.

The \textit{WaterRing} geometry has a 0.5-m-thick water volume around a PC-filled SSS segmented by spherical, rigid unsupported vessels. The water volume around the Sphere is a ring truncated at the poles, to avoid complications with interpolating the temperatures at higher/lower regions than the approximate measuring heights of the temperature probes. A separate initialization was used for the water and SSS' interior, using the following linear interpolation:

\begin{equation}
T_{N/S}(t,y) = \frac{1}{h_1^i-h_0^i} \big( T_{N/S}^{i+1}(t) (y-h_0^i) + T_{N/S}^i(t) (h_1^i-y) \big)
\label{eq:boundary_t}
\end{equation}

where $h^i_{0/1}$ is the interpolation domain's upper(lower) height limit, which obviously coincides with the lower(upper) limit for the next $i+1$ (previous $i-1$) domain; $y$ is the vertical coordinate in any of the model's internal points, and $T_{N/S}^i$ are the recorded (North or South side) temperatures used as anchors on the square domains' corners, taken from the historical, time-varying LTPS Phase I(.a or .b) dataset. Care was exercised in order to keep the vertical coordinate of the LTPS probes the same as the domains' vertical limits, even though the horizontal position of the anchor point may vary slightly ($\mathcal{O}(10cm)$) to ensure smooth physical interpolation within the internal region of interest. 

The fact that this model is closed and enables fluid movement requires careful handling of the simulating conditions, in particular to the iterative timing, mesh geometry and iteration divergence probability.
 
To ensure an appropriate numerical modeling the mesh for the external part of the WaterRing benchmark has been meshed using a radial Cartesian grid, with local refinement close to the boundaries representing the internal buffers. The internal part, representing the IV, has been meshed using both Cartesian and polygonal meshes in order to check accuracy and independence of numerical results from it. Such analysis showed a dependence for the Cartesian grid, with preferential direction of the fluid inside the IV, and the total independence with the use of the polygonal mesh. Based on such analysis and the guidelines carried out from the benchmarking cases reported in Section\ref{sec:Numerical modeling} and \ref{sec:phenobench} the typical cell size $(\Delta x)$ is around $3\cdotp10^{-2}$ m, the number of cells $1.2\cdotp10^{5}$ and the time-step size$\Delta t$ equal to $9$s. This enabled for the best compromise between computational efficiency and expediency, and appropriate low numerical noise levels as described above.
 
A custom-made software tool took care of checking, at each iteration, at which point the simulated time was, comparing it with the listed times in the recorded data from the LTPS probes. Once this simulated time reached or exceeded a given time limit (set at 1800 s, since that is the standard time delay between data acquisitions by the LTPS sensors), the appropriate historical temperatures were updated as imposed boundary conditions on the water ring's surface. Provisions were implemented to ensure good data would always be available: in case of dropouts in DAQ, the imposed temperature would be kept at the last available time. This can cause a slight upset once new data is available, but the need to select relatively small periods for computation efficiency meant the dropout periods were short and few --furthermore, these jumps were verified not to cause large enough deviations on the boundary conditions to motivate divergences or unphysical effects on the numerical solver. Conditions inside the SSS were left free too evolve when $t\neq0$, including on its boundary. This smoothed out the ring-to-SSS initialization differences in a few numerical iterations.

This model's benchmarking power was realized by placing "tallying spots" in the nominal positions where the Phase I.a sensors would be. Simulated temperatures on these points, 1 m away from the imposed boundary condition, could be checked against the historical recorded data, to verify good thermal transport behavior across this instrumented region, representative of the whole Inner Detector. These cases focused on $\sim$1-month periods during the \textit{Transient} and \textit{Insulated} periods described in \cite{previous_paper}. The earliest possible date for this model would be April 10th, 2015, since the Phase I.b sensors entered operation then. This roughly coincides with the end of the \textit{Transient} period.

\paragraph{Results} Figure~\ref{fig:residuals_WR_ins} shows the residuals ("true" historical temperature minus simulated temperature at the same time and position) for the 14 positions of the LTPS Phase I.a probes. Good agreement can be seen, with a smooth exponential trend toward stable errors, up to $\sim <$2.25$^{\circ}$C --although equilibrium errors are no bigger than $\sim$0.15$^{\circ}$C. Discretization and temperature interpolation account for this level of errors. The overall trend shows a remarkable agreement between recorded and simulated data, as shown in Figure~\ref{fig:residuals_WR_ins}. Even more remarkably, the simulation shows a situation whereupon the temperature initialization profile that sent temperatures to slightly incorrect values, due to the interpolation strategy followed, is corrected by the time evolution profile and allows the behavior to follow the recorded data profile within ~2 days of simulated time. It should be noted the 1-1.3 m/day thermal inertia in the detector\cite{other_paper} accounts for at least part of this delay.

\begin{figure}[tb]
\centering 
\includegraphics[width=1\columnwidth]{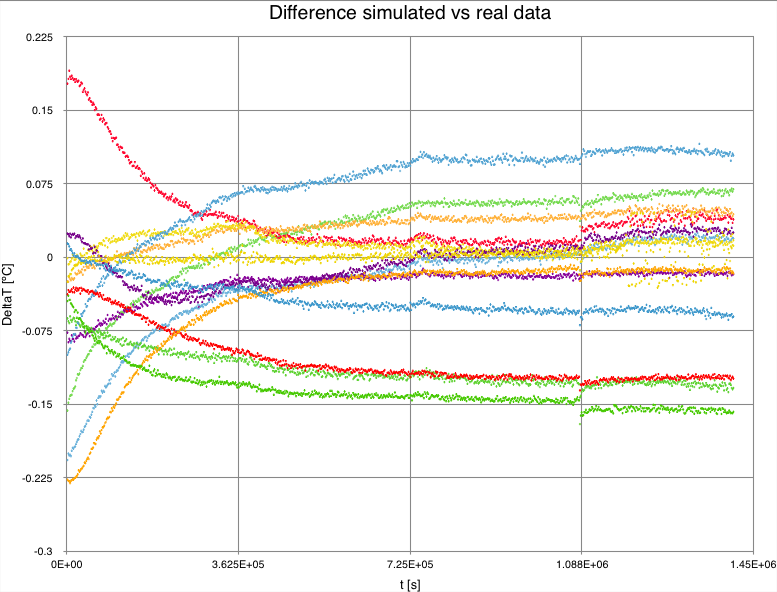} 
\caption{Residuals between real recorded temperatures and simulated ones for the same locations in the same timeframe (insulated period).}
\label{fig:residuals_WR_ins} 
\end{figure}

A temporal phase shift is evident when focusing on the sharpest available features in the recorded temperatures and their simulated counterpart, as depicted, for example, in Figure~\ref{fig:temps_WR_trans_detail}. This shift is constant and the features (if the phase shift is cancelled out manually) are seen to line up almost perfectly, albeit with a certain --small-- decrease in slope change. The cause for this effect is still under investigation, but is considered not to negatively impact the overall reliability of the thermal transport benchmarking power of this model, given it represents a small and constant shift, and is moreover shown not to cause feature broadening.

\begin{figure}[tb]
\centering 
\includegraphics[width=1\columnwidth]{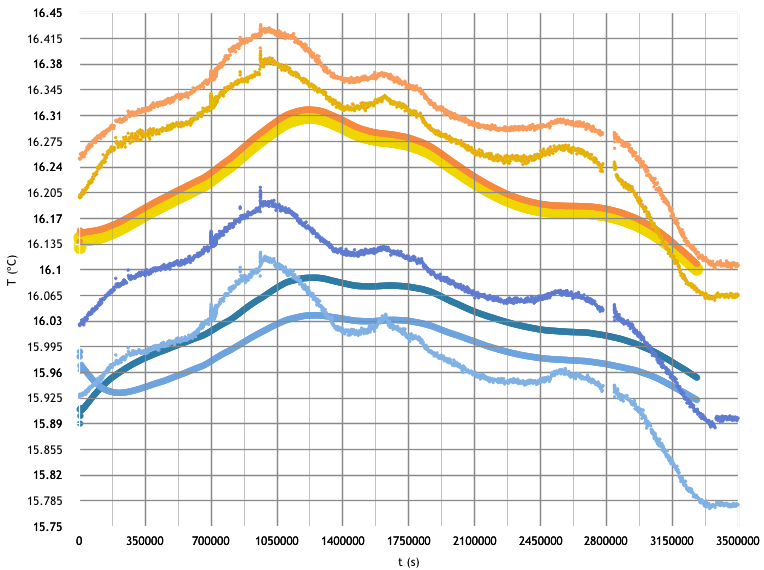} 
\caption{Detail of the real (jagged lines) and simulated (smoother lines) temperature profiles for the transient time period considered, for the two top sensors (67$^{\circ}$ and 50$^{\circ}$). A phase shift of $\sim$1-2 days is visible, apart from the obvious slight offset (maximal for these four sensors) in the abscises axis.}
\label{fig:temps_WR_trans_detail} 
\end{figure}

The \textit{Water Ring} models are seen to provide a powerful benchmark for thermal transport, quite faithfully ($< \pm$0.2$^{\circ}$C, and much better in some cases) replicating the temperature evolution in the OB's LTPS Phase I.a probes positions when the boundary condition represented by the Phase I.b sensors is imposed $\sim$1 m away, in a different medium (water) and having to pass through the SSS structural element. Therefore, at least as far as thermal transport capabilities, the implemented \textit{FLUENT} models are a useful tool to understand, replicate and foresee the thermal environment in the detector. Further, it is reasonable to believe that this extrapolation will hold, for the same geometry (and possibly for similar ones), at other points in the model. 

It is not, however, a benchmark for \textit{fluid} transport: there is no "ground truth" data from Borexino, since no "tracer" is available --other than the same backgrounds we are trying to study through this research.

\section{Fluid dynamics in the ROI (SSS$\rightarrow$IV)}
\label{sec:SSS_IV}

The WaterRing benchmarking model was adapted to include just the Inner Detector, neglecting the water around it, and imposing the Phase I.a buffer probes readings on the SSS boundary, assuming a constant temperature from the real position of the LTPS sensors to the SSS surface, at the same vertical position. An increase of node density (with several cycles of mesh adaptation to improve performance and reliability) and a finer timestep of 4.5 simulated seconds per iteration was thus possible. CFD-derived temperatures were recorded at several points on the Inner Vessel in order to impose them as boundary conditions later on. Effectively, the benchmarking power of the \textit{Water Ring} models was employed as a confidence anchor to ensure the temperatures at the vessel boundary, where no probe is available, would be accurate. These measures should allow for minimization of systematic model-dependent errors, which could not be shown to be uncorrelated enough with the \textit{Water Ring} mesh geometries.

\subsection{IV-only}

\paragraph{Setup} The IV is modeled as a perfect circle of nominal Inner Vessel radius of 4.25 m. No attempt at modeling the actual vessel shape is made, although the deviations are small enough ($\leq \pm$20 cm, or $\sim$0.2$\%$) for this to be a good approximation. The polygonal mesh approach is used, with a cell size of $\sim$5 cm$^2$ ($\mathcal{O}(10^5)$ cells. No internal structures or localized mesh tightening is employed away from the model's boundary. Initialization is performed picking the simulated temperatures a few centimeters outside the vessel in the \textit{Water Ring} models. This small distance away from the vessel is chosen so as to avoid boundary layer effects that may locally shift the isotherms in a way that would falsify the most realistic temperature mapping in the bulk of the IV. As such, these temperatures were also used as input for a time-evolution script, in order to impose time-varying boundary conditions on the model's outer wall.

\paragraph{Boundary conditions} A perfectly-stratified model was first run in order to characterize the level of unphysical currents induced by the numerical iterative process, yielding a background level of $\sim \mathcal{O}(10^{-5})$ m/s. It is noted part of these currents, despite having a physical origin (especially in the horizontal direction), are mesh-enhanced. The absence of boundary currents along the vessel is notable, in sharp contrast with the model with time-changing realistic temperatures imposed as boundary conditions. The model's intrinsic numerical noise level, if the mesh can be regularized, can be much higher ($\mathcal{O}(10^{-8}-10^{-9})$ m/s), although the circular geometry of this model prevents a mesh that is completely regular all over the geometry. Trial runs with a so-called "quad" mesh, with four approximately-checkerboard patterns that converge in a central rectangular mesh, akin but not equal to the very first meshes employed in these simulations, yielded these order-of-magnitude currents, but induced localized unphysical instabilities in the transition areas that would be unassumable in a realistic case. A three-dimensional model may be able to sidestep these geometrical instabilities, but the large computational time required left this potential for numerical noise reduction as a future perspective to be developed.

Five time-dependent models were then run, using the temperatures derived from as many other \textit{Water Ring} cases. These were chosen from the Phase I.b dataset in order to characterize the model in the most diverse array of thermal profiles as possible, from mid-2015 to the latest data available in late winter 2017. This time span comprises the end of the "Transient" period\cite{other_paper} and four regimes during the "Insulated" period, including the phase with the warm, stable top, and the later phase with a relatively-rapidly cooling top. The simulated time periods were between 2 and 8 weeks long.

\begin{figure}[tb]
\centering 
\includegraphics[width=1\columnwidth]{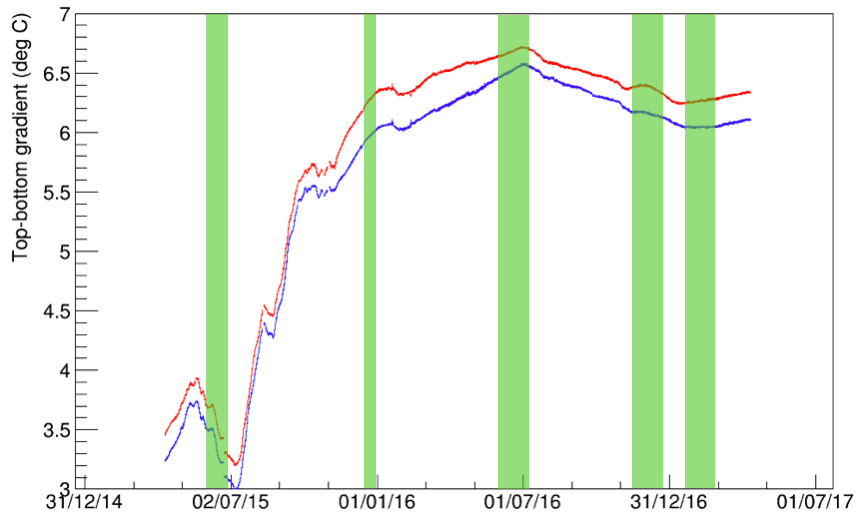} 
\caption{Time periods used in the IV simulations, superimposed on a plot of the (North or South, blue or red, respectively) vertical temperature gradient between the topmost and bottommost Phase I.b probes in the water around the SSS.}
\label{fig:wr_periods} 
\end{figure}

\subsection{Results} 

Horizontal currents are the dominant feature in all of the convective models considered with Borexino geometries and data-based temperature fields. Indeed, as expectable from the Rayleigh number calculation in Section~\ref{sec:methodology}, the large energy gap between the top and bottom, separating the stably-stratified fluid layers, precludes bulk, organized motion in the IV. The spherical geometry, in contrast to the cylindrical symmetry from the cases in Section~\ref{sec:phenobench}, will have a role on this --but asymmetries between both sides are the main driving force behind this preponderance of horizontal currents in place of vertical ones in realistic Borexino scenarios. Indeed, these asymmetries favor elongated recirculation cells that transport fluid from one side of the Sphere to the other one, while leaving the stratification in place, when they favor the breaking of organized vertical movement into minimal-energy transport along slightly-inclined isotherms.

This is better observed through iso-stream-function views, which show the discretized local fluid-carrying capacity of the flow (see Figure~\ref{fig:stft}), and allow for a better identification of the general fluid behavior, rather than studying the current velocities directly. These may skew the attention toward small high-speed regions (which can be very localized and have no global importance, or even be numerically-induced by mesh irregularities): while these regions may propel fluid, and with it, background radioisotopes, at the highest velocities, they need not account for most of the background injection into the FV --indeed, often they are not correlated. Current velocities may be useful for determining organized global movement (i.e. in case there was a vertical global trend, as in the cylindrical benchmarks), of which we see no evidence in these cases. Tracking virtual massless particles "attached to the fluid" through the \textit{pathline} Fluent utility provides confirmation of this fact.

\begin{figure}[tb]
\centering 
\includegraphics[width=1\columnwidth]{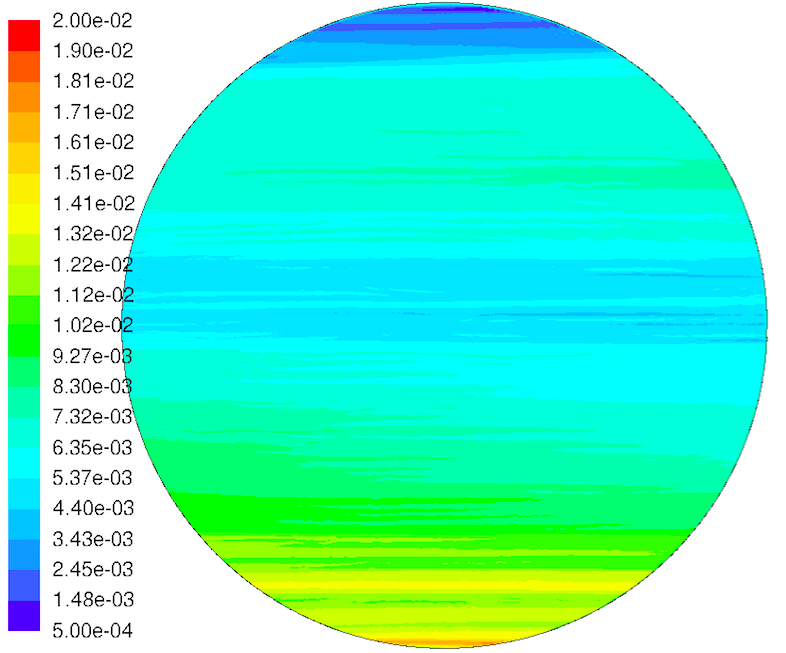} 
\caption{Typical iso-stream-function instantaneous map (in kg/s) for an IV-only model with time-varying boundary conditions based on realistic temperatures, showing fluid carrying capacity of the flow, and therefore indicating where the areas of maximum bulk transport are located.}
\label{fig:stft} 
\end{figure}

Furthermore, a common characteristic to all models is a strong (up to 5 orders of magnitude) surface current along the vessel internal periphery (see Figure~\ref{fig:IV_lower_current}), although in general it does not span the whole surface with the same direction, owing to the inhomogeneous vertical separation between isotherms. It is worthwhile to note that this feature was not present in any area of the stably-stratified control model, indicating it is a feature uniquely derived from the boundary conditions and geometry, but definitely not numerically-induced. This feature was hypothesized as an explanation for the introduction of background components (particularly $^{210}$Po) from the less-radiopure vessel nylon into the FV's scintillator, although its detachment mechanism was the subject of much speculation and remained unexplained before numerical simulations were carried out.

\begin{figure}[tb]
\centering 
\includegraphics[width=0.8\columnwidth]{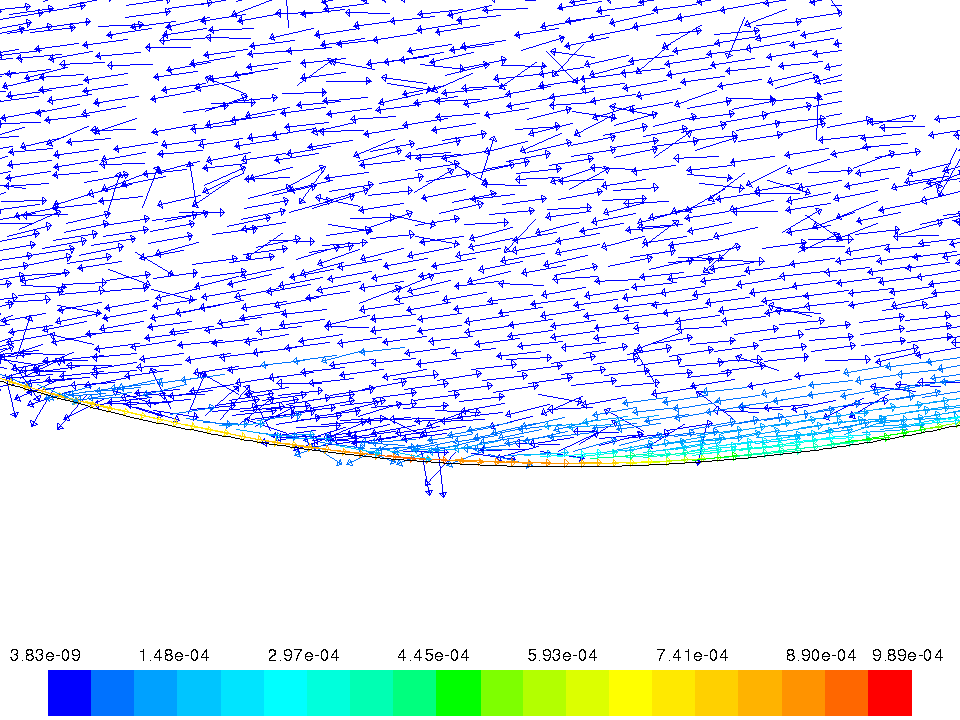} 
\caption{Strong currents along the IV-only models with time-varying boundary conditions based on the numerical propagation of LTPS Phase I.b temperature probe data to the Inner Vessel surface. This feature is prominent and common to all IV models, except the control, stably-stratified one with adiabatic boundary conditions, showing its model independence.}
\label{fig:IV_lower_current} 
\end{figure}

Interestingly, this rules out the idea of a "chimney" effect picking up radioimpurities from the vessel and propelling them up or down the vertical axis of the vessel. For similar temperature fields, and changes, as the ones studied here for the 2015-17 timespan, the dominant mixing mechanism is instead mainly horizontal. Extended discussion of background stability and transport with fluid movement will not be further discussed here, being out of the scope of the current CFD work. Instead, it will be reported in an upcoming publication detailing the techniques used for low-statistics background tracking through data selection and localization, and its correlation with CFD scenarios under development.

Horizontal velocities between $\mathcal{O}$(10$^{-5}$) and $\mathcal{O}$(10$^{-7}$) m/s were seen to be the largest in magnitude in all models, including the control scenario with ideally stratified bulk with adiabatic boundary conditions. Even though these currents are seen to be mesh-enhanced locally, thanks to the aforementioned control stratified adiabatic case, large scale features are thought to be physical, since a markedly different horizontal current distribution is seen in the realistic scenarios, and much less horizontal organization develops in the stratified case, as well as with slower currents.

No large-scale organized vertical motion was seen to exist, and areas of vertical velocity larger than mesh cell hotspots (still, no larger than a few tens of centimeters) are typically between $\mathcal{O}$(10$^{-7}$) and $\mathcal{O}$(10$^{-9}$) m/s. This discussion excludes the boundary layer in close proximity to the vessel boundary, where current magnitudes are much larger (up to $\sim$10$^{-3}$ m/s in some cases).

\section{Conclusions and prospects}
\label{sec:conc}

The current work provides valuable insights into the determination of non-turbulent convective dynamics in a closed, pseudo-stably stratified system such as Borexino, showing excellent benchmarking reproducibility in the Rayleigh number ranges of 10$^5$-10$^7$, and appropriate large-scale reproducibility down to $\mathcal{O}$(10$^3$). Attainment of global, organized convective modes is seen to exist in cylindrical geometries with no threshold in the time of $\Delta T$ application, its lateral symmetry or its magnitude, as long as this $\Delta T$ is applied on the lateral walls. Conversely, only local convection will appear, up to the vertical distance the isotherms will be displaced, when applying the $\Delta T$ to just the terminal caps. In a spherical geometry such as the Borexino detector, good thermal reproducibility was reached by comparing the large historical dataset of recorded temperatures to the temperature field obtained in CFD runs at the same positions. This allowed for the study of the smallest unsegmented region of interest possible, the Inner Volume of the detector, where the understanding of small currents causing the mixing of the scintillator it contains is critical to further improve radioactive background levels that may allow the detector data to be of even higher quality. Horizontal movement caused by lateral asymmetrical imbalances in the boundary conditions is seen to be the main driving factor in bringing fluid from the periphery to the center of the volume, while no global forced convective fields are seen to develop. Observed horizontal currents are seen to be of the order and span that would be of concern for background transport ($> \mathcal{O}$(10$^{-7}$), but this is not so for vertical currents. Carrying capacity maps ("iso-stream-function plots") clearly mark the regions where most of the fluid motion occurs, offering a powerful tool to understanding past behavior in the detector, as well as to engineer minimal-mixing temperature profiles for future directives toward establishing an ultra-low level of scintillator mixing --and, with it, unprecedented levels of radioactive background presence in Borexino's Fiducial Volume.

The present results may not only inform the particular case of the Borexino neutrino observatory's internal facilities, but also expand the limited modeling of non-turbulent fluid mixing in pseudo-steady-state closed systems near equilibrium, subject to small asymmetrical perturbations in their temperature field, such as liquid reservoirs (water, liquified gas, petroleum-derived, deep cryogenics...), which share equivalent geometries and conditions.

\section{Acknowledgements}

The Borexino program is made possible by funding from INFN (Italy), NSF (USA), BMBF, DFG, HGF and MPG
(Germany), RFBR (Grants 16-02-01026 A, 15-02-02117 A, 16-29-13014 ofi-m, 17-02-00305 A) (Russia), and NCN
Poland (Grant No. UMO-2013/10/E/ST2/00180). We acknowledge the generous hospitality and support of the Laboratori Nazionali del Gran Sasso (Italy). All numerical simulations reported in this study are performed in the HPC system of the interdepartmental laboratory 'CFDHub' of Politecnico di Milano.

\section*{References}
\footnotesize
\bibliography{Paper_Borexino_CFD}
\normalsize
\end{nolinenumbers}
\end{document}